# The geographic proximity effect on domestic cross-sector vis-à-vis intra-sector research collaborations

Giovanni Abramo[1], Francesca Apponi[2] and Ciriaco Andrea D'Angelo[3]

[1] *giovanni.abramo@iasi.cnr.it*
Laboratory for Studies in Research Evaluation, Institute for System Analysis and Computer Science (IASI-CNR), National Research Council of Italy (Italy)

[2] *francesca.apponi@uniroma2.it*
Department of Engineering and Management, University of Rome "Tor Vergata" (Italy)

[3] *dangelo@dii.uniroma2.it*
Department of Engineering and Management, University of Rome "Tor Vergata" (Italy)
&
Laboratory for Studies in Research Evaluation, Institute for System Analysis and Computer Science (IASI-CNR), National Research Council of Italy (Italy)

**Abstract**
Geographic proximity is acknowledged to be a key factor in research collaborations. Specifically, it can work as a possible substitute for institutional proximity. The present study investigates the relevance of the "proximity" effect for different types of national research collaborations. We apply a bibliometric approach based on the Italian 2010-2017 scientific production indexed in the Web of Science. On such dataset, we apply statistical tools for analyzing if and to what extent geographical distance between co-authors in the byline of a publication varies across collaboration types, scientific disciplines, and along time. Results can inform policies aimed at effectively stimulating cross-sector collaborations, and also bear direct practical implications for research performance assessments.

# 1. Introduction

Policies in developed countries are aimed at improving efficiency both in knowledge creation and knowledge diffusion among actors involved in different stages of the knowledge value chain. The "triple helix" model (Etzkowitz & Leydesdorff, 1998) envisages the close interaction among the public research area, industrial system, and government institutions as the best way to enhance innovation and development of nations. In this regard, university-industry collaboration allows the match between knowledge, sources, and competence of two realities to achieve higher impact in the long term (Briggs 2015; Briggs & Wade 2014; Su, Lin, & Chen, 2015). Public-private research collaboration is also one of the main channels which favors knowledge transfer, because it achieves both knowledge creation and transfer at once (D'Este & Patel, 2007).

Among various elements that have an influence on the phenomena of knowledge transfer and research collaboration, proximity is generally acknowledged to be a key factor (Boschma, 2005). Proximity is described as "being close to something measured on a certain dimension" (Knoben & Oerlemans, 2006). It facilitates coordination, enables communication among actors and reduces uncertainty. Specifically, the probability of collaborations is shaped by five types of proximities: organizational, institutional, geographical, social, and cognitive (Boschma 2005). The importance of proximity dimensions varies across different types of interaction (Alpaydin & Fitjar, 2020), and some forms of proximity might compensate the effect of others.

For instance, geographic proximity can work as a possible substitute for institutional proximity (Crescenzi, Filippetti, & Iammarino, 2017). In cross-sector collaborations, university and industry have to face the institutional differences that could influence their interaction. The closeness among partners might facilitate cross-sector collaboration, with the geographic proximity compensating for the absence of other proximities, since it increases the possibility of personal interaction and the transfer of tacit knowledge.

The purpose of this work is to investigate the geographic proximity effect on cross-sector collaborations and contrast it with intra-sector collaborations taking place among researchers belonging to the same sector. Previous studies on the geographic distance between inter-institutional collaborations have investigated either type of collaborations alone, but never at once, that is observing the same actors, in the same environment and time period.

In this study, we compare the relevance of the proximity effect for three types of collaboration: public-public, public-private and private-private. We further distinguish between "national only" and "international also" collaborations, to understand if the presence of an international partner might influence the average distance between domestic partners. We also distinguish among the different scientific disciplines, since the intensity of public-private collaboration varies across research fields (Abramo, Apponi & D'Angelo, 2021). Finally, we investigate whether the geographic proximity effect varies along time, as it occurs in the case of knowledge spillovers, where it decays over time (Abramo, D'Angelo & Di Costa, 2020a).

In particular, we intend to answer the following research questions:
- How does the proximity effect impact public-private research collaborations?
- Are there any differences with respect to intra sector public-public or private-private collaborations?
- Does the proximity effect vary in the presence of international collaborations, across fields, and along time?

In order to answer these questions, we conduct a large-scale analysis on the Italian 2010-2017 scientific production indexed in Web of Science (WoS). We measure the average distance of all pairs of authors in the by-line of over 335,000 Italian publications, and apply some statistical tools for analyzing the relationship between distance and type of collaboration across scientific disciplines. The reason why the analysis is restricted to Italy only, is that reconciling and disambiguating private sector affiliations as distinct from public ones in foreign countries is a formidable task for non nationals.

Results could inform policies aimed at stimulating cross-sector interaction. They also bear direct practical implications for research performance assessments along the cross-sector collaboration dimension, which might need to account for the geographic proximity effect, not to disfavor relatively remote institutions (Abramo, D'Angelo, & Solazzi, 2011). We warn the reader that these kinds of studies are, by nature, inevitably domestic in scope, as the geography of the country and localization of organizations therein heavily affect results. Consequently, we recommend caution in generalizing results, or even comparing them with those of other national contexts.

We exploit the past years' inroad of bibliometrics that makes it now possible to expand the scope and period of observation of investigations on the topic. Specific bibliometric methodologies were developed specifically for this purpose (Abramo, D'Angelo, & Solazzi, 2011). However, the advantages come together with an observation bias that the reader should be aware of. As this methodology is based on research publication output, it captures only successful collaborations (otherwise the work would not be published). Moreover, not all co-authored publications reveal a real collaboration, and not all successful collaboration necessarily lead to publications.

In the next section, we review the literature on the influence of the proximity effect on private-public research collaboration. In Section 3, we present the methodology and data. In Section 4, we show the results of the analysis and, in the last section, we conclude the study with our consideration.

## 2. Literature review

A large body of the literature has investigated the particular characteristics that could influence the effectiveness of cross-sector research collaboration: size, sector, and R&D intensity of firms; size and scientific specialization of universities (Spithoven, Vlegels & Ysebaert, 2019).

A number of investigations on various countries, scope, methods and indicators on the spatial distance between public and private organizations engaging in research collaboration have been conducted (Giuliani & Arza, 2009; Hewitt-Dundas & Roper, 2011; Tijssen, Waltman, & Van Eck, 2011; Autant-Bernard, Billand, & Massard, 2012; Giunta, Pericoli & Pierucci, 2016). A trend noticed by many is that the average distance between partners becomes wider over time (Waltman, Tijssen, & Van Eck, 2011; Alpaydın, 2019; Abramo, D'Angelo & Di Costa, 2020b), which reflects the globalization of research.

The influence of geographic proximity on the effectiveness of knowledge transfer, as well as a catalyzer of research collaboration, has been demonstrated in cases when the tacit component of knowledge to be shared is conspicuous. When knowledge is transferable mainly through demonstration and observation, requiring face-to-face interaction among partners, knowledge transfer is more easily achieved if the actors are co-located (Gertler, 2003; Morgan 2004; Singh, 2005), or when the geographic proximity between partners allots frequent interactions (Hong & Su, 2013; Garcia, Araujo, Mascarini, Gomes Santos & Costa, 2015; Fitjar & Gjelsvik, 2018; Petruzzelli & Murgia, 2020).

Universities tend to collaborate with industries located within a limited geographical distance because of lower coordination costs, higher effectiveness in face-to-face interactions, and based on the existence of a common context (Fitjar & Gjelsvik, 2018). However, it appears easier to overcome geographic than cognitive distance (Arant, Fornahl, Grashof, Hesse & Söllner, 2019). In fact, when partners are cognitively close, they tend to interact at larger geographical distances (Garcia, Araujo, Mascarini, Gomes Dos Santos & Costa, 2018).

Moreover, the interplay of geographical distance and quality of the university partners also influences both collaborations and outcomes. Both geographical proximity and research quality appear positively associated with the frequency of university-industry partnerships; however, differences occur across scientific disciplines (D'Este & Iammarino, 2010). The geographic proximity of university and industry might favor research collaboration even if there is a trade-off between the quality of local universities and the higher costs associated with greater distance (Guerrero, 2020; Tang, Motohashi, Hu & Montoro-Sanchez, 2020). In the UK, physical co-location with top-tier universities favors cross-sector collaboration. However, if faced with this choice, UK firms (especially the R&D-intensive ones), appear to prefer quality over distance (Laursen, Reichstein & Salter, 2011). It was shown an inverted-U shape relationship between excellence of university partners and distance with industry partners (D'Este & Iammarino, 2010). In a subsequent study, the same authors found that firms located in intensive R&D clusters tend to partner with universities regardless of their location, while firms outside such clusters tend to partner with local universities (D'Este, Guy, & Iammarino, 2013).

On the contrary, Abramo, D'Angelo and Solazzi (2011) revealed a problem of information asymmetry in the market for university-industry research collaboration in Italy. The authors found that, in 93% of cases, firms could have collaborated with a higher quality university. In 54% of cases, there was at least one university both closer and of higher quality when compared to the university that was actually chosen for collaboration. At single professor level, in 95% of cases the private company could have partnered with a higher performing professor in the same field of the collaboration; in 65% of cases, the choice could have been a better-performing professor, located closer to the company. Tijssen, van de Klippe, & Yegros (2020) identified a number of determinants affecting university-industry research collaborations, varying across distance zones. Four of them appear to be common to all zones, namely intensive R&D firms, research size of a university and its quality, and gatekeepers among the faculty.

## 3. Data and methods

In order to answer the research questions, an econometric analysis of the Italian scientific production of the period 2010-2017 was carried out. The data source is the Italian National Citation Report, extracted from the Web of Science *core collection* imposing "Italy" as affiliation country of at least one author. The unit of observation is the single scientific publication resulting from national extramural collaboration. In WoS, each bibliometric address is composed of two parts: the first one refers to the affiliation and is made up of four "segments", corresponding in general to the macro-organization (Seg1) and to its internal articulations at the level of "School" (Seg2), "Department" (Seg3) and "Research unit" (Seg4). The second part consists of toponymic information: City, Province, State, Zip_Code and Country. Therefore, in order for a publication to be defined as the result of national extramural collaboration, the following two conditions referring to the byline must be met:
- It must contain at least two authors;
- It must contain at least two Italy addresses referring to distinct organizations, i.e., distinct "Seg1 - City" pairs.

We measure the geographic proximity between co-authors of a publication in terms of the average geodesic distance of the cities associated with all distinct Seg1+City pairs found in the publication's address list.[1] This distance is a function of geographical coordinates of cities extracted from the Italian institute of statistics (ISTAT)[2] for Italian LAUs.[3] For reasons of computational complexity, publications with more than ten distinct "Seg1 - City" pairs are excluded,[4] as well as those in which we are unable to geo-locate all the cities indicated in the address list, due to a transcription error in the source data. In total, the analysis dataset includes 335,574 publications. As an example, we report the case of the publication with accession number WOS:000208151600003, whose address list is given in Table 1.

*Table 1: The address list of a publication in the dataset*

| Seg1 | Seg2 | Seg3 | Seg4 | CITY | PROVINCE | STATE | ZIP_CODE | Country |
|---|---|---|---|---|---|---|---|---|
| Univ Pavia | Dept Appl Hlth Sci | Sect Med Stat & Epidemiol | | Pavia | | | I-27100 | IT |
| Univ Pavia | Dept Math | | | Pavia | | | I-27100 | IT |
| Ca Granda Osp Maggiore Policlinico | Serv Biostat | | | Milan | | | | IT |
| Osped Niguarda | Serv Biostat | | | Milan | | | | IT |

This list consists of three distinct Seg1 - City pairs
- Univ Pavia - Pavia
- Ca Granda Osp Maggiore Policlinico - Milan
- Osped Niguarda - Milan

As these are public research organizations (one university and two hospitals), the publication is classified as "national intra-sector public extramural collaboration". In case one (or more) public and one (or more) private national organization(s) are recognizable in the address list, the publication is classified as "cross-sector national collaboration". Disambiguation of public vs private organizations requires manual scrutiny and profound knowledge of the country under observation. A subsequent step is reconciliation of all bibliographic addresses with "Italy" as affiliation country (D'Angelo, Giuffrida, & Abramo, 2011). Through such reconciliation it is possible to tag a publication as fruit of:

- Intra-sector public collaboration: if all "Seg1 - City" pairs related to the Italian addresses, pertain to recognized public national organizations;
- Intra-sector private collaboration: if all "Seg1 - City" pairs, pertain to recognized private Italian organizations;
- Cross-sector collaboration: if "Seg1 - City" Italian pairs in the address list pertain both to public and to private national organizations.

---

[1] Authors with multiple affiliations are counted multiple times. This would entail a distortion in the average value of the distance computed for publications by such authors. However, in Italy multiple affiliations hardly concern national institutions, rather, a national institution and a foreign one, and we do not observe the latter.
[2] https://www.istat.it/it/archivio/6789, last accessed on 2 November 2021.
[3] The LAU level consists of municipalities or equivalent units. The assumption of the "city" as an element of geo-referencing implies that any collaboration between organizations located in the same city occurs at zero distance.
[4] These amount to 1.3% of total observations. Even considering that this elimination affects some disciplines more than others, the effect remains very limited in size.

Finally, the possible presence of one or more addresses with a country different from "Italy" implies the international tagging of the publication.

It is understood that the presence/absence of a foreign address does not affect the average value of the distance between the authors of a publication, which is exclusively calculated with reference to the Italian addresses, being the aim of the work the analysis of the geographical proximity in the national extramural collaborations.

In order to deepen the analysis at field level, each publication in the dataset is assigned to the subject category of the hosting journal.[5] Considering the aggregation of subject categories in macro-areas[6], Table 2 shows the breakdown of the 335,574 publications in the dataset by type and macro-area.

*Table 2: Analysis dataset by type of publication and macro-area*

|  | Intra-sector public (national) | Intra-sector public (international) | Cross-sector (national) | Cross-sector (international) | Intra-sector private (national) | Intra-sector private (international) | Total |
|---|---|---|---|---|---|---|---|
| Mathematics | 5,724 | 3,296 | 198 | 87 | 2 | 1 | 9,308 |
| Physics | 22,442 | 32,142 | 2,151 | 803 | 41 | 29 | 57,608 |
| Chemistry | 14,245 | 10,207 | 1,246 | 506 | 20 | 10 | 26,234 |
| Earth&Space science | 12,484 | 10,494 | 1,161 | 527 | 16 | 12 | 24,694 |
| Biology | 31,297 | 19,894 | 1,849 | 1,002 | 26 | 34 | 54,102 |
| Biomedical Research | 37,119 | 20,669 | 1,414 | 891 | 22 | 24 | 60,139 |
| Clinical Medicine | 69,385 | 37,044 | 1,678 | 906 | 11 | 22 | 109,046 |
| Psychology | 2,567 | 1,729 | 34 | 17 | 0 | 0 | 4,347 |
| Engineering | 37,084 | 23,191 | 5,977 | 2,078 | 143 | 106 | 68,579 |
| Economics | 4,381 | 2,708 | 204 | 120 | 9 | 3 | 7,425 |
| Law, political&social science | 4,475 | 2,543 | 157 | 67 | 2 | 2 | 7,246 |
| Art and Humanities | 1,832 | 757 | 106 | 43 | 1 | 0 | 2,739 |
| Multidisciplinary | 623 | 553 | 25 | 32 | 3 | 2 | 1,238 |
| Total* | 189,632 | 128,438 | 11,940 | 5,157 | 217 | 190 | 335,574 |

*\* The total is less than the sum of column data, due to multiple counting of publications in SCs falling in more than one macro-area.*

---

[5] In the case of multi-category journals, the publication is assigned indiscriminately to all subject categories (multiple counting).

[6] Our assignment of SCs to macro-areas follows a pattern previously published on the website of ISI Journal Citation Reports, but no longer available on the current Clarivate portal.

An OLS regression was performed on the dataset for estimating how the average geographic distance between co-authors of a publication varies as a function of:
- Type of collaboration, specified by 2 dummies: Cross-sector ($X_1$); Intra-sector private-private ($X_2$), considering Intra-sector public-public as baseline.
- Presence of at least one foreign affiliation, specified by a dummy variable ($X_3$).
- Number of authors in the byline ($X_4$).
- Presence of at least one university, specified by a dummy variable ($X_5$).

In order to estimate a possible temporal pattern in the data, we consider an additional dummy ($X_6$), assuming 1 for 2014-2017 publications and 0 for 2010-2013 ones. Finally, in order to control for area effects, we also consider 13 additional dummies, one for each macro-area.

## 4. Results

We will initially present the descriptive analysis of geographic distances in the different types of collaboration and at the macro-area level. Next, we will illustrate the results of the inferential analysis.

### 4.1 Descriptive analysis

The distribution of the average distance values between the co-authors of publications resulting from different types of collaboration, as shown in Figure 1, reveals already at a glance the presence of a differentiated proximity effect for cross-sector versus intra-sector collaborations. The former have higher values of both central tendency (mean and median) and interquartile distance. Given the geography of the country, the maximum values (all around just over 1,000 km) obviously tend to saturate. Table 3 reports the full descriptive statistics of the average distance for the publications in the dataset. For each analyzed set, the high skewness determines a very significant deviation of the mean values from the medians. In particular, intra-sector public national collaborations show an average distance of 132.7 km and a median distance of 47.1 km. Intra-sector private national collaborations have longer distances, with an average of 147.3 km and a median of 50.3 km. The figure for cross-sector national collaborations rises further to 148.2 km and 80.4 km for their mean and median, respectively. The presence of a foreign partner seems to show a significant effect on the mean/median distance of private intra-sector and cross-sector collaborations, but no effect on public intra-sectors.

It should also be noted that at least a quarter of the total number of publications is the result of collaboration between researchers located in the same city, even if in different organizations, as indicated by the value of the 25th percentile of the distribution of distances, invariably zero for all the sets under analysis, except for cross-sector international collaborations (where the average distance is less than 8 km). Regarding the variability, the distributions of intra-sector public and cross-sector collaborations show very similar values of standard deviation, around 180-190 km. In contrast, private intra-sectors show significantly greater variability, with a standard deviation of 210 km for nationals and 220 km for internationals. Regarding the international dimension of the collaboration, it seems to show a significant effect on the average distance of Italian co-authors of a private intra-sector and cross-sector publication. In contrast, in the absence of a private organization in the byline, the presence of a foreigner does not appear to be associated with a change in mean/median distance between the co-authors of a publication.

*Figure 1: Box plot of average distance of publications' co-authors, by collaboration type*

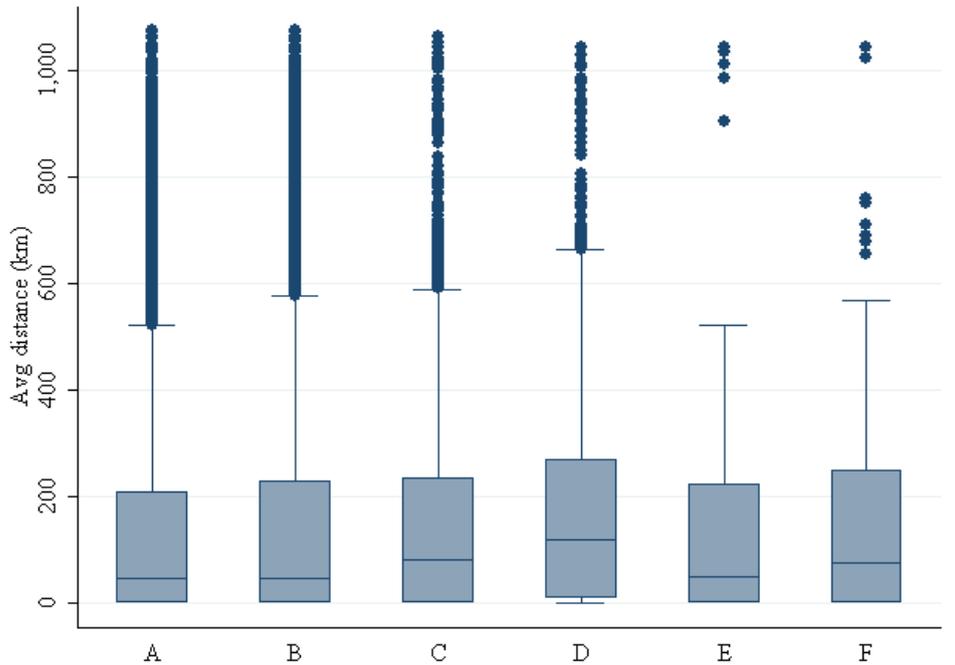

A = Intra-sector public national; B = Intra-sector public international; C = Cross-sector national; D = Cross-sector international; E = Intra-sector private national; F = Intra-sector private international

*Table 3: Descriptive statistics of distances (in km) between publications' co-authors, by collaboration type*

|  | Intra-sector public national | Intra-sector public international | Cross-sector national | Cross-sector international | Intra-sector private national | Intra-sector private international |
|---|---|---|---|---|---|---|
| Obs | 189,632 | 128,438 | 11,940 | 5,157 | 217 | 190 |
| Mean | 132.7 | 136.6 | 148.2 | 171.3 | 147.3 | 166.1 |
| Std Dev. | 184.1 | 183.8 | 178.4 | 191.0 | 210.3 | 219.8 |
| Max | 1,074.9 | 1,074.9 | 1,063.8 | 1,042.8 | 1,042.8 | 1,045.0 |
| Skewness | 1.780 | 1.596 | 1.631 | 1.504 | 2.071 | 1.669 |
| Kurtosis | 6.291 | 5.549 | 6.058 | 5.676 | 7.881 | 5.844 |
| 25% | 0 | 0 | 0 | 7.8 | 0 | 0 |
| 50% | 47.1 | 45.6 | 80.4 | 118.8 | 50.3 | 74.4 |
| 75% | 209.2 | 230.5 | 236.1 | 270.1 | 222.8 | 249.3 |
| 90% | 396.0 | 402.8 | 392.5 | 442.7 | 468.6 | 477.4 |
| 95% | 513.6 | 512.0 | 502.0 | 528.0 | 511.7 | 567.7 |

The following figures show the breakdown of the data by macro-area. Figure 2 shows the average distances for publications that are the result of intra-sector public collaborations, and it can be seen that, for the national case, the greatest distances are in Mathematics, Social Sciences (in particular Economics) and Art and Humanities. The shortest distances are in Clinical Medicine and Biomedical Research. In these two areas, the presence of at least one author with a foreign affiliation significantly increases the average distance between partners. This is also occurring in Psychology, while the opposite is true in all other areas. Cross-sector collaborations are a different matter: Figure 3 shows that, for the national case, maximum average distances are found in Clinical Medicine and Biomedical Research, while minimum ones are found in Mathematics and Law, Political and Social Sciences. The presence of authors with foreign affiliation increases the average distance between cross-sector partners in all areas but three: Law, Political and Social Sciences, Psychology, Biomedical Research.

*Figure 2: Average distances (in km) between co-authors of intra-sector public publications by macro-area*

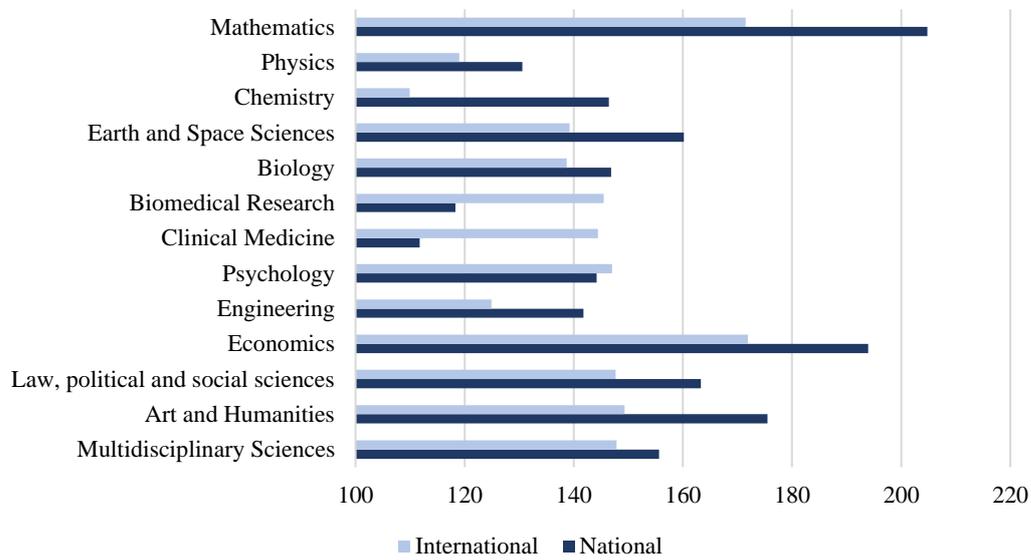

*Figure 3: Average distances (in km) between co-authors of cross-sector publications by macro-area*

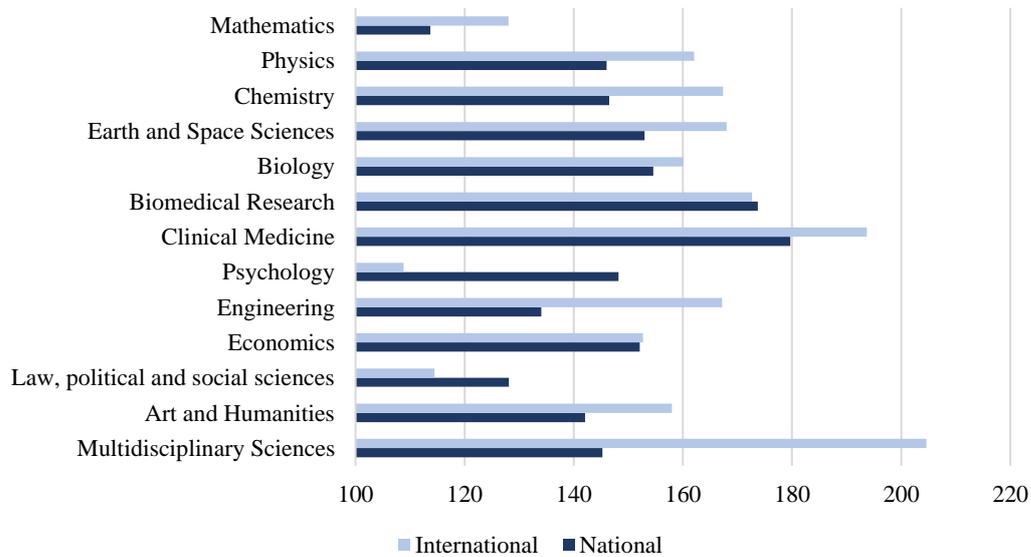

## 4.2 Inferential analysis

In order to answer the research questions, the OLS regression described in Section 3 was implemented. Table 4 reports the descriptive statistics of the model variables and Table 5 the correlation indexes between pairs of variables.

The fourth column of Table 4 indicates that publications resulting from cross-sector collaborations represent just over 5% of the total. Those with at least one foreign author are just under 40%. 84.6% of the publications have at least one address attributable to a university. The average number of authors of the publications of the dataset is 8.2 and, finally, we notice a slight imbalance of the dataset on the second period: the publications of the four-year period 2014-2017 are 56.4% of the total, against 45.6% of 2010-2013.

*Table 4: Average values of the regression model variables*

| Variable | | | Mean | Std Dev. | Max | Median | Skewness | Kurtosis |
|---|---|---|---|---|---|---|---|---|
| Response | Y | Avg distance | 135 | 184 | 1075 | 48 | 1.699 | 2.983 |
| | $X_1$ | Cross-sector | 0.051 | 0.220 | 1 | 0 | 4.084 | 14.7 |
| | $X_2$ | Intra-sector private | 0.001 | 0.035 | 1 | 0 | 28.7 | 820 |
| Collaboration | $X_3$ | International | 0.399 | 0.490 | 1 | 0 | 0.414 | -1.829 |
| | $X_4$ | No of authors | 8.2 | 19.2 | 3,036 | 6 | 49.4 | 5,748 |
| | $X_5$ | Presence of universities | 0.846 | 0.361 | 1 | 1 | -1.913 | 1.661 |
| Other | $X_6$ | Period | 0.564 | 0.496 | 1 | 1 | -0.256 | -1.934 |

Table 5 reveals a practically non-existent correlation between the variables of the model, which leads us to exclude possible multicollinearity effects: the highest coefficient (0.140) concerns the $X_3$-$X_4$ pair, indicating a weak association between the international character of the publication and the number of its authors.

*Table 5: Correlation matrix of the variables of the regression model*

| | Y | $X_1$ | $X_2$ | $X_3$ | $X_4$ | $X_5$ | $X_6$ |
|---|---|---|---|---|---|---|---|
| Y | 1 | | | | | | |
| $X_1$ | 0.025 | 1 | | | | | |
| $X_2$ | 0.004 | -0.008 | 1 | | | | |
| $X_3$ | 0.012 | -0.046 | 0.005 | 1 | | | |
| $X_4$ | 0.063 | -0.008 | -0.003 | 0.140 | 1 | | |
| $X_5$ | 0.104 | -0.008 | -0.071 | -0.100 | 0.013 | 1 | |
| $X_6$ | 0.028 | 0.037 | 0.003 | 0.044 | 0.013 | 0.021 | 1 |

*Y, Avg distance; X1, Cross-sector; X2, Intra-sector private; X3, International; X4, No. of authors; X5, Presence of universities; X6, Period*

Table 6 shows the results of the OLS regression. On the left side, the data related to the overall analysis (Model 1) are indicated, while on the right side are those obtained considering area effects (Model 2). The coefficients are all positive and significant.

Other things being equal and compared to publications resulting from collaboration between all and only researchers from public organizations, those resulting from cross-sector collaborations show a higher average distance (coeff. 21.85, in model 1). Intra-sector private ones also show greater distances than intra-sector public ones (coeff. 61.62).

The average distance between authors in an extramural collaborative publication increases with the number of co-authors, albeit rather narrowly, i.e., less than one km for each additional author in the byline. The distance also increases with the presence of foreign co-authors (coeff. 5.14).

Even the presence of an academic, other things being equal, seems to have a significant effect of increasing the average distance between the co-authors of the publications, a distance that tends to increase over time as shown by the positive and significant coefficient of the variable period.

The last column of Table 6 shows that the model betas do not vary significantly when area effects are considered, so all the results and effects highlighted with the specification of model 1 are repeated with that of model 2.

With regard specifically to area effects, it is noted that compared to the baseline (Physics) the average distances of national extramural collaborations are greater in all other areas, except for Clinical Medicine. In marginal terms, among the covariates, intra-sector private seems to have the most important effect. With other parameters equal and considering also the area effects, compared to an intra-sector public publication, the intra-sector private collaboration is characterized by a higher average distance of almost 60 km, while the cross-sector collaboration average distance is higher by almost 20 km.

*Table 6: OLS regression, dependent variable: average distance of publications' co-authors (Model 2 embeds area effects with "Physics" as baseline)*

|  | Model 1 | | | Model 2 | | |
| --- | --- | --- | --- | --- | --- | --- |
|  | Coeff. | Std Err. | t | Coeff. | Std Err. | t |
| _cons | 77.42*** | 0.937 | 82.65 | 64.44*** | 1.239 | 52.02 |
| Cross-sector | 21.85*** | 1.436 | 15.22 | 19.35*** | 1.261 | 15.34 |
| Intra-sector private | 61.62*** | 9.076 | 6.79 | 58.65*** | 9.162 | 6.40 |
| International | 5.14*** | 0.655 | 7.85 | 3.24*** | 0.690 | 4.70 |
| Number of authors | 0.57*** | 0.017 | 34.69 | 0.71*** | 0.082 | 8.69 |
| Presence of universities | 53.29*** | 0.879 | 60.60 | 55.82*** | 0.734 | 76.02 |
| Period | 8.77*** | 0.637 | 13.77 | 7.59*** | 0.559 | 13.57 |
| Art and Humanities |  |  |  | 45.99*** | 3.886 | 11.83 |
| Biology |  |  |  | 20.58*** | 1.102 | 18.68 |
| Biomedical Research |  |  |  | 4.12*** | 1.042 | 3.95 |
| Chemistry |  |  |  | 9.56*** | 1.403 | 6.82 |
| Clinical Medicine |  |  |  | -0.40 | 0.934 | -0.43 |
| Earth&Space science |  |  |  | 34.12*** | 1.435 | 23.77 |
| Economics |  |  |  | 58.93*** | 2.504 | 23.54 |
| Engineering |  |  |  | 14.27*** | 1.106 | 12.90 |
| Law, polit.&social scie. |  |  |  | 33.40*** | 2.383 | 14.01 |
| Mathematics |  |  |  | 65.07*** | 2.487 | 26.16 |
| Multidisciplinary |  |  |  | 27.81*** | 5.487 | 5.07 |
| Psychology |  |  |  | 18.45*** | 2.831 | 6.52 |
| Number of obs | 335,574 | | | 432,705 | | |
| df | 6 | | | 18 | | |
| F | 917.9 | | | 577.1 | | |
| Prob > F | 0.000 | | | 0.000 | | |
| R-squared | 0.016 | | | 0.023 | | |
| Root MSE | 182.5 | | | 181.9 | | |

*Statistical significance: \*p-value <0.10, \*\*p-value <0.05, \*\*\*p-value <0.01*

The proximity effect thus seems much more relevant in collaborations between researchers in public organizations. In cross-sector collaborations, average distances between partners increase. They also grow in the presence of foreign authors and authors belonging to universities. All of the above have clear sectoral specificities.

The very low values of R-squares indicate the importance of considering more control variables, especially other dimensions of proximities which bibliometric metadata can hardly capture.

## 5. Discussion and conclusions

Public-private research collaborations are one of the most relevant targets of developed countries policies aiming at improving efficiency both in knowledge creation and knowledge diffusion. Understanding the ways in which they are implemented, the motivations behind for involved partners and the factors that hinder them is key to optimizing such policies. Among elements that have an influence on research collaboration, proximity is generally acknowledged to be a key factor. In cross-sector collaborations, the partners belong to different worlds and have to face institutional differences that could heavily influence their interaction. The cultural, motivational and linguistic "distance" between a researcher working in a university or a public research institution and a colleague working in a private company probably makes it more necessary to resort to face-to-face interactions for the development of the necessary trust, the set up and the tuning of optimal conditions for the achievement of the aims of the collaboration. In such conditions, geographic proximity can work as a possible substitute for institutional proximity. This is at least what we get from scanning previous literature on the subject.

In this study, we tried to verify the presence of this "compensation" effect between geographical and institutional proximity, referring to the Italian context. Our results indicate that this effect is not detectable; on the contrary, in cross-sector collaborations the average distances between partners are greater than in collaborations involving partners from the same sector, i.e., institutionally more similar. One could hypothesize the presence of an "intermediation" effect of quality of the prospect partner. In particular, from the perspective of private firms, especially the R&D-intensive ones, they could prefer quality over distance. This evidence, which has emerged in several studies related to the UK context, has, however, already been refuted for the Italian case. Abramo, D'Angelo and Solazzi (2011) have in fact found the existence of an information asymmetry that would prevent Italian firms, in at least half of the cases, to choose as their partners for possible research collaborations excellent researchers closely located to the company headquarters.

Rather, the result that emerged in the study conducted can find an explanation in the specificity of the context of analysis and, in particular, in the non-homogeneous distribution of private R&D activities on the Italian territory. According to the latest statistical survey just published (ISTAT, 2021), more than 75% of Italian R&D expenditure by private companies is concentrated in five of the twenty regions: apart from Lazio (located in the center of Italy), the remaining four are all Northern regions: Lombardy, Emilia-Romagna, Piedmont and Veneto. The whole of the South covers only just over 9% of national business expenditure. These data attest to an evident greater difficulty for a researcher from a university or public research institution in the South to find a potential industrial partner for a research collaboration of mutual interest, located nearby or within the same region. This has obvious implications for research performance assessments of the so-called "third mission" of universities, an assessment to which in Italy a part of the ordinary funding provided by the Ministry of University and Research is linked.

In contrast, public research activity is less concentrated and more geographically distributed: this would explain why intra-sector public collaborations are characterized by a lower average value of distance between project team members than cross-sector collaborations. This value is also lower than that recorded for intra-sector private collaborations. Basically, the proximity effect is more evident in collaborations between exclusively public researchers than in collaborations between exclusively private firms. However, this last result should be read considering the low number of observations (just over 400) related to this last type of collaboration.

However, for the same type of collaboration, the presence of a university researcher leads to an increase in the average distances between the members of the collaboration team. Evidently, university research is characterized by being less "local", i.e., less constrained by the geographical factor. One of the expected results concerns the effect of the presence of a foreign author in the byline of a publication. When opening up to the international dimension, collaboration between national partners seems to be less affected by the proximity effect: the presence of a foreign colleague in the team makes it less critical to have to interact with colleagues (from the same or another sector) who are not closely located.

A trend emerging from our study, and confirming a number of previous investigations, is that the average distance between partners tends to grow over time under the same conditions and, in particular, independently of the institutional proximity between the partners.

Certainly, our study is by nature inevitably domestic in scope, as the geography of the country heavily affects results, as already widely argued. In particular, in a paper that the authors are about to prepare, they intend to verify if and how much the results obtained depend on the geographical distribution/concentration of R&D activities in Italy. Consequently, we recommend caution in generalizing results, or even comparing them with those of other national contexts. In addition, there are the intrinsic limitations of the bibliometric approach: i) observing publication's authorships allows to capture only successful collaborations; ii) not all co-authored publications reveal a real collaboration, and not all successful collaborations necessarily lead to publications. Nevertheless, the authors believe that these limitations are largely counterbalanced by the power of the approach itself in terms of numerousness, a power found in the high level of significance of the analyses conducted on the phenomenon of interest.